\documentclass[aps,prb,twocolumn,superscriptaddress]{revtex4}
\usepackage{graphicx}% Include figure files
\usepackage{dcolumn}% Align table columns on decimal point
\usepackage{hhline}

\newcolumntype{d}[1]{D{.}{\cdot}{#1}}
\newcolumntype{.}{D{.}{.}{-1}}
\newcolumntype{,}{D{,}{,}{-1}}
\begin{document}

% Use the \preprint command to place your local institutional report
% number in the upper righthand corner of the title page in preprint mode.
% Multiple \preprint commands are allowed.
% Use the 'preprintnumbers' class option to override journal defaults
% to display numbers if necessary
%\preprint{}

%Title of paper
\title{
Film-thickness dependence of 
10~GHz Nb coplanar-waveguide resonators
}

% repeat the \author .. \affiliation  etc. as needed
% \email, \thanks, \homepage, \altaffiliation all apply to the current
% author. Explanatory text should go in the []'s, actual e-mail
% address or url should go in the {}'s for \email and \homepage.
% Please use the appropriate macro foreach each type of information

% \affiliation command applies to all authors since the last
% \affiliation command. The \affiliation command should follow the
% other information
% \affiliation can be followed by \email, \homepage, \thanks as well.
\author{Kunihiro Inomata}
\email{k-inomata@zp.jp.nec.com}
\affiliation{RIKEN Advanced Science Institute, 
34 Miyukigaoka, Tsukuba, Ibaraki 305-8501, Japan 
%Wako-shi, Saitama 351-0198, Japan
} 
\author{Tsuyoshi Yamamoto}
\affiliation{RIKEN Advanced Science Institute, 
34 Miyukigaoka, Tsukuba, Ibaraki 305-8501, Japan 
%Wako-shi, Saitama 351-0198, Japan
} 
\affiliation{NEC Nano Electronics Research Labs., %Laboratories, 
34 Miyukigaoka, Tsukuba, Ibaraki 305-8501, Japan}
%\affiliation{CREST-JST, Kawaguchi, Saitama 332-0012, Japan} 
%
\author{Michio Watanabe}
\altaffiliation{Present address: Fort Lupton Fire Protection District, 
   1121 Denver Avenue, Fort Lupton, Colorado 80621, U.S.A.}
\affiliation{RIKEN Advanced Science Institute, 
34 Miyukigaoka, Tsukuba, Ibaraki 305-8501, Japan 
%Wako-shi, Saitama 351-0198, Japan
} 
\author{Kazuaki Matsuba}
%\affiliation{CREST-JST, Kawaguchi, Saitama 332-0012, Japan} 
\affiliation{%Department of Materials Sciences and Engineering, 
Tokyo Institute of Technology, 4259 Nagatsuta-cho, 
Midori-ku, Yokohama, 226-8503, Japan}
\author{Jaw-Shen Tsai}
\affiliation{RIKEN Advanced Science Institute, 
34 Miyukigaoka, Tsukuba, Ibaraki 305-8501, Japan 
%Wako-shi, Saitama 351-0198, Japan
} 
\affiliation{NEC Nano Electronics Research Labs., %Laboratories, 
34 Miyukigaoka, Tsukuba, Ibaraki 305-8501, Japan}
%\affiliation{CREST-JST, Kawaguchi, Saitama 332-0012, Japan} 
%
%Collaboration name if desired (requires use of superscriptaddress
%option in \documentclass). \noaffiliation is required (may also be
%used with the \author command).
%\collaboration can be followed by \email, \homepage, \thanks as well.
%\collaboration{}
%\noaffiliation

\date{April 14, 2009}
%\date{\today}

\begin{abstract}
We have studied Nb $\lambda/2$ 
coplanar-waveguide (CPW) resonators 
whose resonant frequencies are $10-11$~GHz. 
The resonators have different film thicknesses, 
$t=0.05,$ 0.1, 0.2, and 0.3~$\mu$m. 
We measured at low temperatures, $T=0.02-5$~K, 
one of the scattering-matrix element, $S_{21}$, 
which is the transmission coefficient from one port to the other.  
At the base temperatures, $T=0.02-0.03$~K, 
the resonators are overcoupled to  
the input/output microwave lines, 
and the loaded quality factors are on the order of 
$10^3$. 
The resonant frequency %$f_r$ 
has a considerably larger film-thickness 
dependence compared to the predictions 
by circuit simulators 
which calculate the inductance of CPW 
taking into account $L_g$ only, 
where $L_g$ is the usual magnetic inductance 
determined by the CPW geometry.   
By fitting a theoretical  
$S_{21}$ vs.\ frequency curve 
to the experimental data,  
we determined for each film thickness,  
the phase velocity %$v_p$ 
of the CPW 
with an accuracy better than 0.1\%. 
The large film-thickness dependence must be due to 
% kinetic inductance $L_k$ of the CPW center conductor.  
the kinetic inductance $L_k$ of the CPW center conductor.  
We also measured $S_{21}$ 
as a function of temperature up to $T=4-5$~K, 
and confirmed 
that both thickness and temperature dependence are 
consistent with the theoretical prediction 
for $L_k$. %the kinetic inductance. 
%Based on the theory, we discuss the penetration 
%depth in our Nb films and %superconducting transition temperature 
%its dependences on the film thickness and on temperature. 
\\
\\
J. Vac. Sci. Technol. B {\bfseries 27}, 2286 (2009) 
[DOI: 10.1116/1.3232301]
\end{abstract}

% insert suggested PACS numbers in braces on next line
%\pacs{}

% insert suggested keywords - APS authors don't need to do this
%\keywords{}

%\maketitle must follow title, authors, abstract, \pacs, and \keywords
\maketitle

% body of paper here - Use proper section commands
% References should be done using the \cite, \ref, and \label commands
%%%%%
%%%%%
\section{Introduction}
\label{sec:intro}
Microwave resonators (for example, 
Chap.~7 of Ref.~\onlinecite{Poz90}) 
are one of the key components 
in a variety of circuits operated at GHz frequencies, 
and their new applications continue to emerge. 
A simple example is band-pass filters, which are based on 
the fact that the microwave transmission through resonators 
is frequency sensitive. 
The same idea is also used for more complex devices, 
such as oscillators, tuned amplifiers, and frequency meters. 
Actually, having high-quality filters and oscillators 
is critical in mobile communications, 
where available bands keep getting 
overcrowded as demand grows rapidly. 

Another application of microwave resonators is 
radiation detectors, 
which often consist of 
sensor heads and read-out circuits, 
and 
resonators can be used  
in the readout circuit.  
When one would like to 
detect at the single-photon level, 
one needs to have detectors with 
high enough energy resolutions. 
In this respect, superconducting 
sensor heads\cite{Irw95,Pea96,Gol01} 
can be advantageous, and may be the only solution 
at present depending on the energy range 
of the object. 
Once one decides to use 
superconducting sensor heads, %detectors, 
it makes sense to fabricate  
the read-out circuit with superconducting 
materials as well. 
Superconducting microwave resonators 
allow one to obtain higher quality factors, 
which are favorable for frequency multiplexing. 
In addition, 
one type of photon detector is designed to 
probe the change in the kinetic inductance of 
superconducting thin-film resonator 
due to the absorbed photons.\cite{Maz02} 
In this device concept, the resonator works 
as a sensor head 
rather than a part of the readout circuit.

Recently, superconducting resonators are used   
for the nondemolition %non-destructive 
readout of superconducting qubits as well.\cite{Wal04}  
Since the demonstration by Wallraff {\it et al.},\cite{Wal04}  
this type of readout scheme has been one of the main topics 
in the field of superconducting qubits, and  
we are also developing a similar readout technique.\cite{Ino09} 
In superconducting resonators, kinetic inductance, 
which is essentially the internal mass of 
the current carriers, plays an important role 
especially when the superconducting film is thin.
In our circuit,\cite{Ino09}  
for example, a Nb $\lambda/4$ coplanar-waveguide 
(CPW) resonator is terminated by an Al dc SQUID, 
and the total thickness of the Al layers is 0.04~$\mu$m.  
In order to avoid a discontinuity at the Al/Nb interface, 
we usually choose the Nb thickness to be 0.05~$\mu$m, 
which is much thinner than a typical 
thickness of $\geq$0.3~$\mu$m for superconducting integrated 
circuits fabricated by the standard photolithographic technology. 
Fabricating circuits with thinner films is actually important 
from the viewpoint of miniaturization as well. 
Therefore, for designing resonators, quantitative understanding 
of the kinetic inductance in the CPW is important. 

There have been a number of reports on kinetic inductance for  
a variety of materials.\cite{Mes69,Rau93,Kis93,Wat94,Gub05,Fru05} 
In general, however, kinetic inductance is indirectly measured 
by assuming a theoretical model, and as a result, 
the uncertainties are relatively large.   
%%%%%
%%%%%
Thus, although kinetic inductance is a well established 
notion and the phenomenon is qualitatively understood, 
the quantitative information is not necessarily sufficient 
from the point of view of applications, especially at 
high frequencies, $>$10~GHz.
%%%%%
%%%%%
When we would like to precisely predict  
the resonant frequency, the best solution would be 
to characterize the actual CPW in a simple circuit. 
Such characterization should also improve the %our 
knowledge of superconducting microwave circuits. 
Very recently, G\"{o}ppl {\it et al.}\cite{Gop08} 
measured a series of Al CPW resonators with 
nominally the same film thickness of 0.2~$\mu$m, 
and investigated the relationship between the loaded quality 
factor at the base temperature of 0.02~K
and the coupling capacitance. 
For this purpose, 
it is justified to neglect kinetic inductance    
because the kinetic inductance should 
be the same in their resonators and estimated\cite{Gop08} 
to be about two orders of magnitude smaller than 
the usual magnetic inductance determined by the CPW geometry. % only. 
%Later, we call this magnetic inductance as geometric inductance.
In this work, on the other hand, we paid close attention 
to the resonant frequency as well, and 
characterized Nb CPW resonators as a function of 
film thickness rather than a function of 
coupling capacitance. 
We also looked at the temperature dependence 
in order to discuss kinetic inductance in detail. 

%%%%%%%%%%%%%%%%%%%%
\section{Experiment}
\label{sec:Ex}
%%%%%%%%%%%%%%%%%%%%
%
%%%%%
\begin{table}
\caption
{\label{tab:list}
List of resonators. 
$t$ is the thickness of Nb film; 
$f_r$ is the resonant frequency;  
$Q_L$ is the unloaded quality factor;  
$C_c$ is the coupling capacitance; 
$v_p$ is the phase velocity, and its ratio 
to the speed of light $c$ is listed in percent. 
The values for $f_r$ and $Q_L$ are obtained 
at the base temperatures. 
$C_c$ and $v_p$ are evaluated by least-squares fitting  
(see Fig.~\ref{fig:S21}) with $C=1.6\times10^{-10}$~F/m, 
and their uncertainties are determined by changing the value  
of $C$ by $\pm10\%$, where $C$ is the capacitance 
per unit length.}
\begin{ruledtabular}
\begin{tabular}{cccccc}
 %name 
Reso-	&$t$&$f_r$&$Q_L$&$C_c$&$v_p/c$\\
nator &($\mu$m)&(GHz)&($\times$10$^3$)&(fF)&(\%)\\
 \hline
A1& 0.05           & 10.01 & 1.6& 7.0$\pm$0.4 & 39.31$\pm$0.03 \\
A2& 0.1\phantom{0} & 10.50 & 1.4& 7.3$\pm$0.4 & 41.28$\pm$0.04 \\
A3& 0.2\phantom{0} & 10.74 & 1.4& 7.2$\pm$0.4 & 42.21$\pm$0.04 \\
A4& 0.3\phantom{0} & 10.88 & 1.6& 6.6$\pm$0.4 & 42.71$\pm$0.03 \\
B1& 0.05           & 10.06 & 3.4& 4.6$\pm$0.3 & 39.32$\pm$0.02 \\
B2& 0.1\phantom{0} & 10.56 & 3.1& 4.8$\pm$0.3 & 41.26$\pm$0.02 \\
B3& 0.2\phantom{0} & 10.81 & 2.7& 5.0$\pm$0.3 & 42.27$\pm$0.03 \\
B4& 0.3\phantom{0} & 10.94 & 3.3& 4.5$\pm$0.3 & 42.72$\pm$0.02 \\
\end{tabular}
\end{ruledtabular}
\end{table}
%%%%%
%
%%%%%%%%%%%%%%
%%%%%%%%%%%%%%
%%%%%%%%%%%%%%
\begin{figure}
\begin{center}
\includegraphics[width=0.8\columnwidth,clip]{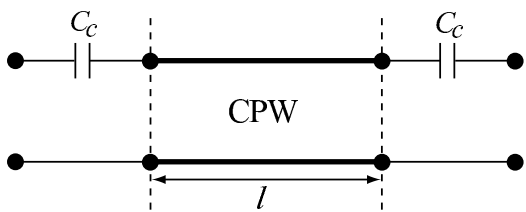}
\caption{\label{fig:CPW_equ}
Schematic diagram of coplanar-waveguide (CPW) resonators. 
A CPW of length $l$ is coupled to the microwave lines 
through capacitors $C_c$. 
}
\end{center}
\end{figure}
%%%
%
%%%%%%%%%%%%%%
%%%%%%%%%%%%%%
%%%%%%%%%%%%%%
\begin{figure}
\begin{center}
\includegraphics[width=0.9\columnwidth,clip]{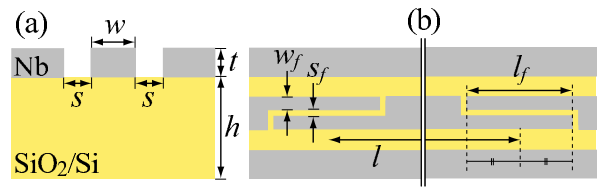}
\caption{\label{fig:CPWshape}
(Color online) 
(a) Cross section of a coplanar waveguide. 
(b) Top view of coupling capacitors. 
}
\end{center}
\end{figure}
%%%
%
We studied two series of Nb $\lambda/2$ CPW resonators 
listed in Table~\ref{tab:list}.  
Each resonator consists of a section of CPW 
and coupling capacitors,  
as shown schematically  
in Fig.~\ref{fig:CPW_equ}. 
The resonators were fabricated on 
a nominally undoped Si wafer whose surface had been 
thermally oxidized. 
On the SiO$_2$/Si substrate, a Nb film was deposited 
by sputtering and then patterned by photolithography 
and SF$_6$ reactive ion etching. 
Figure~\ref{fig:CPWshape}(a) represents the cross section of CPW. 
The center conductor has a width of $w=10$~$\mu$m, 
and separated from the the ground planes by  
$s=5.8$~$\mu$m, so that 
the characteristic impedance becomes $\sim50$~$\Omega$.  
The thickness of Nb is $t=0.05$, 0.1, 0.2, or 
0.3~$\mu$m (see Table~\ref{tab:list}), and 
that of SiO$_2$/Si substrate is $h=300$~$\mu$m.
The SiO$_2$ layer, whose thickness is 0.3~$\mu$m, is 
not drawn in Fig.~\ref{fig:CPWshape}(a).  
We employed interdigital %finger-shaped 
coupling capacitors as shown 
in Fig.~\ref{fig:CPWshape}(b). 
The finger width is $w_f = 9$~$\mu$m, the space between the 
fingers is $s_f = 2$~$\mu$m, and the finger length is 
$l_f = 78$~$\mu$m for Resonators~A1--A4 and 
$l_f = 38$~$\mu$m for Resonators~B1--B4. 
Here, we quoted designed dimensions for the Nb structures. 
The actual dimensions %should 
differ by about 0.2~$\mu$m due to over-etching;  
for example, $w$ and $w_f$ are %expected to be 
$\sim0.2$~$\mu$m smaller, whereas  
$s$ and $s_f$ are $\sim0.2$~$\mu$m larger.      
In this paper, we define the resonator length $l$ as the 
distance between the center of the fingers on one side and 
that on the other side, and $l=5.8$~mm for all resonators.  
Because our chip size is 2.5~mm by 5.0~mm, our CPWs 
meander as in Fig.~\ref{fig:Outline}.  

%%%%%%%%%%%%%%
%%%%%%%%%%%%%%
%%%%%%%%%%%%%%
\begin{figure}
\begin{center}
\includegraphics[width=0.7\columnwidth,clip]{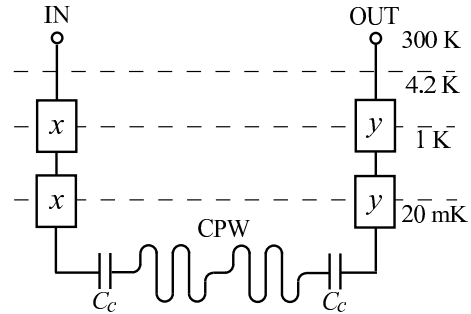}
\caption{\label{fig:Outline}
Schematic diagram of a typical measurement setup. 
Boxes represent attenuators.  
}
\end{center}
\end{figure}
%%%
%

The resonators were measured in 
a $^3$He-$^4$He dilution refrigerator at $T=0.02-5$~K.
A typical measurement setup is shown schematically 
in Fig.~\ref{fig:Outline}. 
The boxes in the figure represent attenuators. 
The amount of attenuation was not the same 
because the microwave lines in our refrigerator 
%(eight in total at the time of measurements) 
had been designed for several different purposes. 
The attenuation was $x = 10$~dB for Resonators~A2, B3, and B4, 
and $x = 20$~dB for the others;  
$y = 10$~dB for all resonators except A1 and A3. 
For Resonators~A1 and A3, we used a line 
with no attenuators ($y=0$~dB) but with 
an isolator and a cryogenic amplifier at 4.2~K. 
The gain of the cryogenic amplifier was 40~dB for 
Resonator~A1 and 34~dB for A3. 
We measured the transmission coefficient 
$S_{21}$ by connecting a vector 
network analyzer to the ``IN" and ``OUT" ports 
in Fig.~\ref{fig:Outline}. 
A typical incident power to the resonator was $-40$~dBm. 
For each resonator, we confirmed that 
the measurements were done in an appropriate power range 
in the sense that the results looked power independent.

%%%%%%%%%%%%%%%%%%%%
\section{Results}
\label{sec:R}
%%%%%%%%%%%%%%%%%%%%
%
%%%%%
\subsection{
%Transmission coefficient 
{\boldmath $S_{21}$} at the base temperatures}
\label{subsec:baseT}
%%%%%
%
%%%%%%%%%%%%%%
%%%%%%%%%%%%%%
%%%%%%%%%%%%%%
\begin{figure}
\begin{center}
\includegraphics[width=\columnwidth,clip]{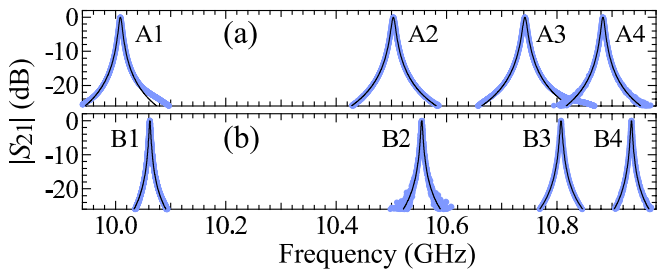}
\caption{\label{fig:S21}
(Color online) 
Amplitude of the transmission coefficient 
$S_{21}$ as a function of frequency for 
(a) Resonators~A1--A4, and (b) Resonators~B1--B4.  
}
\end{center}
\end{figure}
%%%
%

Figure~\ref{fig:S21} shows  
the amplitude of $S_{21}$  
at the base temperatures, $T=0.02-0.03$~K, 
as a function of frequency $f$ 
for all resonators.  
The resonant frequency $f_r$ has a rather large film-thickness 
dependence. Our interpretation is that this is due to 
the kinetic inductance of the CPW center conductor. 
Before discussing the thickness dependence 
%and the kinetic inductance 
in detail, let us look at the quality factors. 

What we obtain by measuring 
%$S_{21}$ vs.\ $f$ curves 
$S_{21}$ as a function of $f$
is the loaded quality factor $Q_L$, which is related to 
the external quality factor $Q_e$ and 
the unloaded quality factor $Q$ by 
\begin{equation}
\label{eq:Q}
Q_L^{-1} = Q_e^{-1}+Q^{-1}.  
\end{equation}
In general, $Q_e$ is determined mainly by $C_c$, 
whereas $Q$ is a measure of the internal loss, 
which arises not only from the dielectric but also 
from the superconductor in the high-frequency regime.  
Our resonators should be highly overcoupled to 
the input/output lines at the base temperatures, 
that is, $Q \gg Q_e$, and thus, $Q_L\sim Q_e$. 
As listed in Table~\ref{tab:list}, $Q_L$ 
of our resonators is on the order of 10$^3$.  
%at the base temperatures. 
These values are not only reasonable for the designs of 
our finger-shaped coupling capacitors 
but also much smaller than typical %quoted 
values of $Q$ below 0.1~K for superconducting 
microwave resonators.\cite{Maz02,Fru05,Gop08} 
When $Q \gg Q_e$, the maximum $|S_{21}|$ is 
expected to be 0~dB. 
We have confirmed by taking into account
attenuators, amplifiers, and cable losses, 
that our measurements are indeed 
consistent within the uncertainties of 
gain/loss calculations, 1--2~dB.
Based on this confirmation, 
the experimental data in Fig.~\ref{fig:S21} 
are normalized so that the peak heights 
equal 0~dB.

The solid curves in Fig.~\ref{fig:S21} are calculations 
based on the transmission ($ABCD$) matrix (for example, 
Sec.~5.5 of Ref.~\onlinecite{Poz90}), and 
they reproduce the experimental data well. 
The matrix for the resonators is given by 
%%%
\begin{equation}
\label{eq:ABCD}
\left(
   \begin{array}{cc}
   A & B \\
   C & D
   \end{array} 
\right)
= T_{\rm cc}\,T_{\rm cpw}\,T_{\rm cc}\,, 
\end{equation}
%%%  
where 
%%%
\begin{equation}
\label{eq:Tcc}
T_{\rm cc} = 
\left(
   \begin{array}{cc}
   1 & (j\omega C_c)^{-1} \\
   0 & 1
   \end{array} 
\right), 
\end{equation}
%%%  
$j$ is the imaginary unit, 
%%%
\begin{equation}
\label{eq:Tcpw}
T_{\rm cpw} = 
\left(
   \begin{array}{cc}
   \cos\beta l & jZ_{\rm cpw}\sin\beta l\\
   j(Z_{\rm cpw})^{-1}\sin\beta l & \cos\beta l
   \end{array} 
\right) 
\end{equation}
%%%  
for lossless CPWs, 
$\omega = 2\pi f$, $\beta = \omega/v_p$, 
%%%
\begin{equation}
\label{eq:vp}
v_p=1/\sqrt{LC}
\end{equation}
%%%  
is the phase velocity, which is strongly related 
to $f_r$, 
%%%
\begin{equation}
\label{eq:Zcpw}
Z_{\rm cpw}=\sqrt{L/C}
\end{equation}
%%%  
is the characteristic impedance, 
$L$ is the inductance per unit length, and  
$C$ is the capacitance per unit length. 
From these transmission-matrix elements, 
the scattering-matrix elements 
are calculated, and $S_{21}$ is given by 
\begin{equation}
\label{eq:S21}
S_{21} = 
2/(A+B/Z_0+CZ_0+D), 
\end{equation}
%%%  
where $Z_0=50$~$\Omega$ is the characteristic impedance 
of the microwave lines connected to the resonator.  
%Unit-length properties of CPW are determined by 
%specifying two parameters out of $v_p$, $Z_{\rm cpw}$, 
%$L$, and $C$. 
Unit-length properties of CPW are determined when  
two parameters out of $v_p$, $Z_{\rm cpw}$, 
$L$, and $C$ are specified. 
In the calculations for Fig.~\ref{fig:S21}, 
we employed $C=1.6\times10^{-10}$~F/m based on the considerations 
described in the following paragraph, and evaluated $C_c$ and 
$v_p$ by least-squares fitting. 

Wen\cite{Wen69} calculated CPW parameters 
using conformal mapping. 
Within the theory, $C$ does not depend on $t$, 
and it is given by 
%%%
\begin{equation}
\label{eq:Wen_C}
C = (\epsilon_r+1)\epsilon_0\,2K(k)/K(k'), 
\end{equation}
%%%  
where $\epsilon_r$ is the relative dielectric constant 
of the substrate, 
$\epsilon_0 = 8.85\times 10^{-12}$~F/m is the 
permittivity of free space,  
$K(k)$ is the complete elliptical integral 
of the first kind, the argument $k$ is given by 
%%%
\begin{equation}
\label{eq:Wen_k}
k = w/(w+2s)\,, 
\end{equation}
%%%  
and $k'= \sqrt{1-k^2}.$ 
For our CPWs, 
we obtain $C=1.6\times10^{-10}$~F/m 
when we employ $\epsilon_r=11.7$ for Si 
(p.~223 of Ref.~\onlinecite{Kit96}) 
neglecting the contribution from the SiO$_2$ layer, 
which is much thinner compared to $w$, $s$, or $h$.  
Circuit simulators [Microwave Office from AWR (\#1) and 
AppCAD from Agilent (\#2)] also predict similar values of $C$.    
%%%%%
%%%%%
The simulators calculate CPW parameters from the 
dimensions and the material used for the substrate. 
%%%%%
%%%%%
The predictions by the simulators have    
$t$ dependence, but in the relevant $t$ range,    
the variations are on the order of 1\% or smaller   
as summarized in Table~\ref{tab:LC},   
and the values of $C$ are between 
$1.6\times10^{-10}$~F/m 
and $1.7\times10^{-10}$~F/m.  
Thus, partly for simplicity,  
we used %in Fig.~\ref{fig:S21} 
$C=1.6\times10^{-10}$~F/m 
for all of our resonators. 

%%%%%
\begin{table}
\caption
{\label{tab:LC}
Dependence of coplanar-waveguide parameters on 
the film thickness $t$. 
For capacitance $C$ and inductance $L$ per unit length, 
the normalized variations 
$\Delta C(t)/C^*$ and $\Delta L(t)/L^*$ 
are listed in percent, where 
$\Delta C(t) = C(t) - C^*$,  
$C^* \sim 1.6 \times 10^{-10}$~F/m  
is the value at $t=0.3$~$\mu$m,   
and the definitions of 
$\Delta L(t)$ and $L^*\sim4\times10^{-7}$~H/m are similar. 
The predictions by circuit simulators \#1 and \#2 
are compared. 
Regarding $L$, 
experimental values 
for ``A"=Resonators~A1--A4 
and for ``B"=Resonators~B1--B4 are also given, 
and they are obtained from the values of $v_p$ in 
Table~\ref{tab:list} using Eq.~(\ref{eq:vp})
and by neglecting the $t$ dependence of $C$.  
}
\begin{ruledtabular}
\begin{tabular}{ccc|cccc}
$t$ & \multicolumn{2}{c|}{$\Delta C(t)/C^*$ (\%)} & 
      \multicolumn{4}{c}{$\Delta L(t)/L^*$ (\%)}\\
($\mu$m) & \#1 & \#2 & \#1 & \#2 & A & B \\
 \hline
0.05           & $-0.6$ & 1.7 & 3.9 & 2.9 & 18.0 & 18.0 \\
0.1\phantom{0} & $-0.5$ & 1.4 & 3.0 & 2.2 & \phantom{1}7.1 
					  & \phantom{1}7.2\\
0.2\phantom{0} & $-0.2$ & 0.8 & 1.4 & 1.2 & \phantom{1}2.4 		                                  & \phantom{1}2.1\\
\end{tabular}
\end{ruledtabular}
\end{table}
%%%%%

With $C=1.6\times10^{-10}$~F/m, the values of $v_p$ 
in Table~\ref{tab:list} correspond to 
$Z_{\rm cpw}=49-53$~$\Omega$, which agrees with 
our design of $\sim50$~$\Omega$. 
We have done the same fitting by changing the value  
of $C$ by $\pm10\%$ as well 
in order to estimate the uncertainties, which are also  
listed in Table~\ref{tab:list}.
Within the uncertainties, the values 
of $C_c$ from the same coupling-capacitor design agree, 
and $C_c\sim7$~fF for Resonators~A1--A4 with $l_f=78$~$\mu$m 
and $C_c\sim5$~fF for Resonators~B1--B4 with $l_f=38$~$\mu$m. 
The uncertainties for $v_p$ is much smaller, $<0.1\%$, 
and again within the uncertainties, the values of $v_p$ 
for the same $t$ agree.

For the rest of this paper, let us 
assume that $t$ dependence of $C$ is negligible. 
This assumption is consistent with the fact that 
the experimental $C_c$ vs.\ $t$ in Table~\ref{tab:list} 
does not show any obvious trend. 
Moreover, according to 
the circuit simulators in Table~\ref{tab:LC},  
$t$ dependence of $C$ is smaller than that of $L$.  
Below, 
we look at $L$ mainly  
instead of $v_p$ or other CPW parameters 
so that we will be able to discuss 
the kinetic inductance. 
As long as we deal with 
a normalized inductance 
such as 
%$L(t)/L(\mbox{0.3~$\mu$m})$,  
the ratio of $L(t)$ to $L^*\equiv L(\mbox{0.3~$\mu$m})$,
what we choose for the value of $C$ does not matter 
very much because $v_p$ obtained from the fitting 
was not so sensitive to $C$.  
Hence, we analyze the quantities 
obtained with $C=1.6\times10^{-10}$~F/m only hereafter. 
In Table~\ref{tab:LC},   
we list the variations of $L$ 
%with respect to $t$ 
in our two series of resonators 
as well. For both series, the magnitude of 
the variations are much larger than the 
predictions by circuit simulators. 
We will discuss this large $t$ dependence in terms of 
kinetic inductance in Sec.~\ref{sec:D} after 
examining the temperature dependence in Sec.~\ref{subsec:Tdep}.

%%%%%
\subsection{Temperature dependence of {\boldmath $S_{21}$}}
\label{subsec:Tdep}
%%%%%

%
%%%%%%%%%%%%%%
%%%%%%%%%%%%%%
%%%%%%%%%%%%%%
\begin{figure}
\begin{center}
\includegraphics[width=0.95\columnwidth,clip]{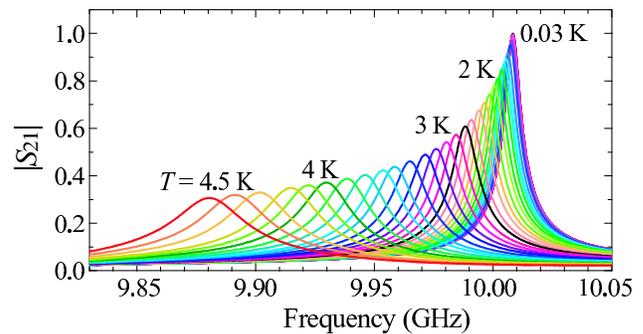}
\caption{\label{fig:S21_temp}
(Color online) 
Amplitude of the transmission coefficient 
$S_{21}$ as a function of frequency for 
Resonators~A1 at different temperatures.  
}
\end{center}
\end{figure}
%%%
%
%%%%%%%%%%%%%%
%%%%%%%%%%%%%%
%%%%%%%%%%%%%%
\begin{figure}
\begin{center}
\includegraphics[width=0.7\columnwidth,clip]{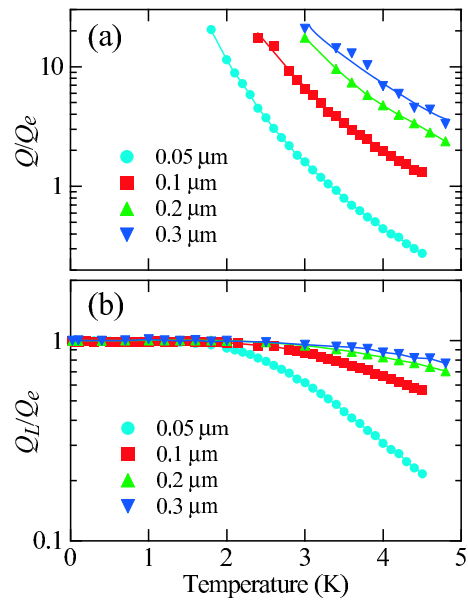}
\caption{\label{fig:Q2}
(Color online) 
Normalized quality factors,  
(a) $Q/Q_e$ and (b) $Q_L/Q_e$, 
as functions of temperature for 
Resonators~A1--A4
(Nb thickness $t=0.05,$ 0.1, 0.2, and 0.3~$\mu$m), 
where $Q_L$, $Q_e$, and $Q$ are 
loaded, external, and unloaded quality factors, respectively, 
and $Q_e$ is assumed to be temperature independent.   
The markers are data points, whereas the curves are 
guides to the eyes. 
}
\end{center}
\end{figure}
%%%
%

We also measured $S_{21}$ vs.\ $f$ 
at various temperatures 
up to $T=4-5$~K for Resonators~A1--A4. 
We show the results %$|S_{21}|$--$f$ curves 
for Resonator~A1 in Fig.~\ref{fig:S21_temp}. 
With increasing temperature, 
$f_r$, $Q_L$, and the peak height 
%maximum $|S_{21}|$  
decrease. 
As in Sec.~\ref{subsec:baseT}, 
let us look at the quality factors first. 
In our resonators, $Q_e\sim Q_L$ 
at the base temperatures as we pointed out  
in Sec.~\ref{subsec:baseT}. 
Thus, when we assume that $Q_e$ is 
temperature independent,  
we can calculate $Q$ 
from measured $Q_L$ 
using Eq.~(\ref{eq:Q}). 
We plot $Q_L(T)/Q_e$ and $Q(T)/Q_e$  
vs.\ $T$ in Fig.~\ref{fig:Q2} 
%as functions of temperature  
for all of the four resonators. 
With increasing temperature, 
$Q$ decreases in all resonators. 
A finite $Q^{-1}$ means that the resonator 
has a finite internal loss, which is 
consistent with a peak height smaller than unity  
in Fig.~\ref{fig:S21_temp}.  
%%%%%
The internal loss at high temperatures must be 
due to quasiparticles in the superconductor, 
as discussed in Ref.~\onlinecite{Maz02}. 
%%%%%
The reduction of quality factors becomes larger 
as the Nb thickness is decreased. 
At $T<1$~K, however, the reduction 
is negligibly small, and thus, in this sense, 
it should be fine to choose  
any thickness in the range of $t=0.05-0.3$~$\mu$m
for the study of superconducting qubits 
that we mentioned in Sec.~\ref{sec:intro} 
because qubit operations are almost 
always done at the base temperatures.

\begin{figure}
\begin{center}
\includegraphics[width=0.7\columnwidth,clip]{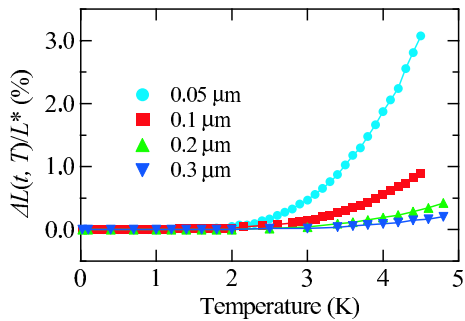}
\caption{\label{fig:L_delta}
(Color online) 
Temperature variations of 
inductance $L$ per unit length  
for Resonators~A1--A4, whose Nb thickness 
is $t=0.05,$ 0.1, 0.2, and 0.3~$\mu$m.  
The unit of the vertical axis is percent. 
See text for the definition of $\Delta L(t,T)/L^*$. 
}
\end{center}
\end{figure}

When CPWs are no longer lossless, $\beta$ 
in Eq.~(\ref{eq:Tcpw}) has to be replaced by 
$(\alpha+j\beta)/j$. %, where $j$ is the imaginary unit. 
This $\alpha$ characterizes 
the internal loss, and $\beta/(2\alpha)$ is 
equal to $Q$ (for example, Sec.~7.2 of 
Ref.~\onlinecite{Poz90}). 
From similar calculations to those 
in Sec.~\ref{subsec:baseT}, we evaluated 
$L$ at higher temperatures as well 
by neglecting the $T$ dependence of $C$ and $C_c$. 
Because we are interested in the temperature variation of $L$, 
we show $\Delta L(t,T)/L^*$ vs.\ $T$ in Fig.~\ref{fig:L_delta}, 
where $\Delta L(t,T) \equiv L(t,T) - L(t,T^*)$,  
$T^*$ is the base temperature, and  
$L^* \equiv L(\mbox{0.3~$\mu$m},T^*)$.  
The variation becomes larger 
as the Nb thickness is decreased.    
This trend also suggests that we should take into account 
the kinetic inductance. 
%%%%%
%of the CPW center conductor. 
%%%%%

%%%%%%%%%%%%%%
%%%%%%%%%%%%%%
%%%%%%%%%%%%%%
%%%
%

%%%%%%%%%%%%%%%%%%%%
\section{Discussion}
\label{sec:D}
%%%%%%%%%%%%%%%%%%%%
%
%%%%%
%\subsection{Kinetic inductance}
%\label{subsec:KI}
%%%%%

The film-thickness and temperature dependence that we have 
examined in %Secs.~\ref{subsec:baseT} and \ref{subsec:Tdep} 
Sec.~\ref{sec:R}
is explained by %within 
the model, 
%%%
\begin{equation}
\label{eq:L}
L(t,T) = L_g(t) + L_k(t,T),  
\end{equation}
%%%  
where  
%$L_g$ is the geometric inductance of the CPW per unit length and  
$L_g$ is the usual magnetic inductance per unit length 
determined by the CPW geometry and  
$L_k$ is the kinetic inductance of the CPW center conductor 
per unit length.  
%%%%%
We neglect the contribution of the ground planes to $L_k$ 
because the ground planes are much wider than the center 
conductor in our resonators [see Eq.~(\ref{eq:Lk})].
%%%%%
We also assume that $L_g$ depends on $t$ only, 
whereas $L_k$ does on both $t$ and $T$.   
This type of model has been employed 
in earlier works\cite{Rau93,Fru05,Gop08} as well. 
The $T$ dependence of $L_k$ arises from the fact that 
%in general, 
$L_k$ is determined not only by the geometry 
but also by the penetration depth $\lambda$, which 
%depends on $T$. 
varies with $T$. 
%Because $\lambda$ depends 
%on $T/T_c$, $L_k$ varies with $T$ dependent. 
%This is the origin of 
Meservey and Tedrow\cite{Mes69} calculated $L_k$ of  
a superconducting strip, and when the strip has a rectangular 
cross section like our CPWs, %in Fig.~\ref{fig:CPWshape}, 
$L_k$ is written as 
%%%
\begin{equation}
\label{eq:Lk}
L_k = \frac{\mu_0}{\,\pi^2\,}(\lambda/w)\ln(4w/t)
\frac{\sinh(t/\lambda)}{\,\cosh(t/\lambda)-1\,}, 
\end{equation}
%%%  
where $\mu_0=4\pi\times10^{-7}$~H/m 
is the permeability of free space. 
The relationship between $L_k$ and $\lambda$ 
is expressed in a much simpler form  
in the thick- and thin-film limits;  
$L_k\propto\lambda$ for 
$t\gg\lambda$,
and 
$L_k\propto\lambda^2$ for 
$t\ll\lambda$. 
When we assume Eqs.~(\ref{eq:L}) and (\ref{eq:Lk}), 
we obtain $\lambda(t,T)$ numerically,    
once $L_g(t)$ is given. 
Below, we 
discuss $\lambda(t,T)$ in our Nb films 
in order to 
confirm that the model represented 
by Eq.~(\ref{eq:L}) is indeed appropriate.

%%%%%%%%%%%%%%
%%%%%%%%%%%%%%
%%%%%%%%%%%%%%
\begin{figure}
\begin{center}
\includegraphics[width=0.8\columnwidth,clip]{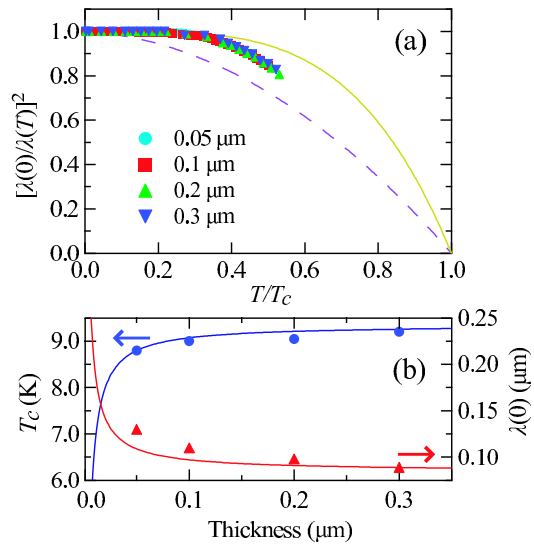}
\caption{\label{fig:lambda_tdep}
(Color online) 
(a) Temperature dependence of the penetration depth $\lambda$ 
for Resonators~A1--A4, whose Nb thickness 
is $t=0.05,$ 0.1, 0.2, and 0.3~$\mu$m. 
The solid and broken curves are theoretical predictions 
expressed by Eqs.~(\ref{eq:2fluid}) and (\ref{eq:Rau}), 
respectively.  
As in Ref.~\onlinecite{Tin96}, 
$[\lambda(t,0)/\lambda(t,T)]^2$ is plotted vs.\ $T/T_c(t)$.    
(b) Superconducting transition temperature $T_c(t)$ used 
in (a), and $\lambda(t,0)$. 
The curves are from Ref.~\onlinecite{Gub05}, 
that is, not fitted to our experimental data.  
Both in (a) and (b), 
$\lambda(T^*)\sim\lambda(0)$ is assumed, where 
$T^*$ is the base temperature.
}
\end{center}
\end{figure}
%%%
%
%%%%%
\begin{table}
\caption
{\label{tab:Lm}
Inductance per unit length 
at the base temperatures 
in Resonators~A1--A4. 
$t$ is the thickness of Nb film; 
$\Delta L_g(t) = L_g(t) - L_g^*$, 
where $L_g$ is the  
usual magnetic inductance per unit length
determined by the CPW geometry, 
and $L_g^*\equiv L_g(\mbox{0.3~$\mu$m})
=3.75\times10^{-7}$~H/m; 
$L_k$ is the kinetic inductance per unit length, 
and $L=L_g+L_k$. 
%Both $\Delta L_m(t)/L_m^*$ and 
%$L_k/L$ are in percent.
}
\begin{ruledtabular}
\begin{tabular}{ccc}
$t$ ($\mu$m) & $\Delta L_g(t)/L_g^*$ (\%)&$L_k/L$ (\%)\\
 \hline
0.05           & 4.3 & 13.1\\
0.1\phantom{0} & 3.4 & \phantom{1}4.9\\
0.2\phantom{0} & 1.7 & \phantom{1}2.2\\
0.3\phantom{0} & -- & \phantom{1}1.6\\

\end{tabular}
\end{ruledtabular}
\end{table}
%%%%%

In Fig.~\ref{fig:lambda_tdep}(a), we plot 
$[\lambda(t,T^*)/\lambda(t,T)]^2$ vs.\ $T/T_c(t)$ 
for Resonators~A1--A4,   
%$[\lambda(0)/\lambda(T)]^2$ vs.\ $T/T_c(t)$ 
%as in Ref.~\onlinecite{Tin96}, 
where $T_c(t)$ 
is the superconducting transition temperature, 
which is assumed to be also $t$ dependent 
in this paper. 
We have found that with a reasonable set of 
parameters, $L_g(t)$ and $T_c(t)$, 
the experimental data for all resonators 
are described by a single curve. 
%%%
%%%
This kind of scaling is expected theoretically 
in the limits of $\xi_0/\lambda_L\gg 1$ and 
$\xi_0/\lambda_L\ll 1$, where $\xi_0$ is the  
coherence length and $\lambda_L$ is 
the London penetration depth.\cite{Tin96} 
Although $\xi_0/\lambda_L\sim1$ in Nb 
(p.~353 of Ref.~\onlinecite{Kit96}) 
at temperatures well below $T_c$, 
it would be still reasonable to expect a scaling 
in our Nb resonators %as well 
because at a given 
normalized temperature $T/T_c(t)$, the relevant 
quantities should be on the same order of magnitude 
in all resonators, and thus, two parameters, 
$\lambda(t,T^*)$ and $T_c(t)$, are probably 
enough for characterizing $\lambda(t,T)$ of 
our resonators. 
%%%
%%%
The values of $L_g(t)$ and $T_c(t)$ employed 
in Fig.~\ref{fig:lambda_tdep}(a) 
are summarized in Table~\ref{tab:Lm} and 
Fig.~\ref{fig:lambda_tdep}(b), respectively.  
The relative change of $L_g(t)$ 
in Table~\ref{tab:Lm} is 
similar to the predictions by circuit simulators 
in Table~\ref{tab:LC}, 
which do not take into account the kinetic inductance. 
The magnitude of $L_g(t)$ is also reasonable because 
$\sqrt{L_g(t)/C}\sim49$~$\Omega$ for all thickness. 
In Table~\ref{tab:Lm}, we also list the ratio of 
kinetic inductance $L_k$ to the total inductance $L$. 
With decreasing thickness, 
$L_k/L$ indeed increases rapidly.  
In Fig.~\ref{fig:lambda_tdep}(b), $T_c(t)$ and 
$\lambda(t,T^*)$ are plotted together with 
the theoretical curves in Figs.~1 and 6 of 
Ref.~\onlinecite{Gub05}, where  
%The curves are partly empirical in the sense that 
Gubin {\it et al.}\cite{Gub05} determined 
some parameters of the curves  
by fitting to their experimental data. 
%We emphasize that the curves are not 
%fitted to our data. 
The values of $T_c(t)$ are reasonable, 
and $\lambda(t,T^*)$ is on the right order of 
magnitude. 

The solid curve in Fig.~\ref{fig:lambda_tdep}(a) 
is the theoretical $T$ dependence 
based on the two-fluid approximation,\cite{Tin96}  
%%%
\begin{equation}
\label{eq:2fluid}
%
%\lambda(T)/\lambda(0) = 1/\sqrt{1-(T/T_c)^4}.  
[\lambda(0)/\lambda(T)]^2 = 1-(T/T_c)^4.  
\end{equation}
%%%  
This theoretical curve reproduces 
the experimental data 
at $T/T_c < 0.4$, when we assume that 
$\lambda(t,T^*)\sim\lambda(t,0)$ 
in Resonators~A1--A4. 
At $T/T_c \geq 0.4$, on the other hand, 
the experimental data deviate from 
Eq.~(\ref{eq:2fluid}), but 
according to Ref.~\onlinecite{Tin96}, 
the expression for $\lambda$ vs.\ $T$ depends on 
the ratio of $\xi_0/\lambda_L$, and thus, 
Eq.~(\ref{eq:2fluid}) %the two-fluid dependence % p. 101 
cannot be expected to apply to all materials equally well.  
%%%
%where $\xi_0$ is the 
%coherence length and  
%$\lambda_L$ is the London penetration depth. 
%%%
Indeed, 
although the temperature dependence of Eq.~(\ref{eq:2fluid})
has been observed in the classic pure superconductors,\cite{Tin96}  
such as Al with $\xi_0/\lambda_L \gg 1$ at temperatures 
well below $T_c$, it does not seem to be the case in 
the high-$T_c$ materials, whose typical $\xi_0/\lambda_L$ 
%is typically $10^{-3}\ll1$ according to Ref.~\onlinecite{Tin96}, 
is in the opposite limit,\cite{Tin96} $\xi_0/\lambda_L \ll 1$,   
and for example,  
Rauch {\it et al.}\cite{Rau93} employed 
for a high-$T_c$ material YBa$_2$Cu$_3$O$_{7-x}$, 
an empirical expression of 
%%%
\begin{equation}
\label{eq:Rau}
%
%\lambda(T)/\lambda(0) = 1/\sqrt{1-0.1(T/T_c)-0.9(T/T_c)^2}.  
[\lambda(0)/\lambda(T)]^2 = 1-0.1(T/T_c)-0.9(T/T_c)^2,    
\end{equation}
%%% 
which is the broken curve in Fig.~\ref{fig:lambda_tdep}(a),  
instead.  
Because $\xi_0/\lambda_L\sim1$ in Nb 
% (p.~353 of Ref.~\onlinecite{Kit96}) 
even at $T/T_c\ll 1$, and because 
the experimental data at $T/T_c \geq 0.4$ 
are between Eqs.~(\ref{eq:2fluid}) and (\ref{eq:Rau}), 
we believe that the deviation from Eq.~(\ref{eq:2fluid}) 
at $T/T_c \geq 0.4$ is reasonable. 

From the discussion in this section, 
we conclude that the model represented 
by Eq.~(\ref{eq:L}) explains 
the film-thickness and temperature dependence of 
our resonators.

%%%%%%%%%%%%%%%%%%%%
\section{Conclusion}
%%%%%%%%%%%%%%%%%%%%
We investigated two series of  
Nb $\lambda/2$ CPW resonators  
with 
resonant frequencies in the range of $10-11$~GHz 
and with 
different Nb-film thicknesses, $0.05-0.3$~$\mu$m.  
We measured the transmission coefficient $S_{21}$ 
as a function of frequency at low temperatures, $T=0.02-5$~K. 
For each film thickness, 
we determined the phase velocity %$v_p$ 
in the CPW with an accuracy better than 0.1\%    
by least-squares fitting of a theoretical  
$S_{21}$ curve based on the transmission matrix 
to the experimental data at the base temperatures.   
Not only the film-thickness dependence but also the 
temperature dependence of the resonators are 
explained by taking into account the kinetic inductance 
of the CPW center conductor.

\section*{Acknowledgment}
The authors would like to thank 
Y. Kitagawa for fabricating the resonators,   
and T. Miyazaki for fruitful discussion. 
T.\ Y., K.\ M., and J.-S.\ T.\ 
would like to thank CREST-JST, Japan 
for financial support. 

%\bibliographystyle{prb}
%\bibliography{KIpaper}

\end{document}